\documentclass[aps, preprint, amsmath, amssymb, prd,
  superscriptaddress]{revtex4-1}

\usepackage{epsfig}
\usepackage{xspace}
\usepackage{multirow}
\usepackage{dcolumn}
\usepackage{braket}
\usepackage{siunitx}

\newcommand{\gev}{~\textrm{GeV}\xspace}
\newcommand{\mev}{~\textrm{MeV}\xspace}

\newcommand{\nb}{~\textrm{nb}\xspace}

\newcommand{\br}{\ensuremath{\mathcal{B}}}

\newcommand{\gcb}{Gaussian\raisebox{.4ex}{\tiny\bf +}Crystal
  Ball\xspace}

\newcommand{\phienu}{\mbox{\ensuremath{D_s \to \phi e \nu}}\xspace}
\newcommand{\etaenu}{\mbox{\ensuremath{D_s \to \eta e \nu}}\xspace}
\newcommand{\etapenu}{\mbox{\ensuremath{D_s \to \eta' e \nu}}\xspace}
\newcommand{\fenu}{\mbox{\ensuremath{D_s \to f_0 e \nu}}\xspace}
\newcommand{\ksenu}{\mbox{\ensuremath{D_s \to K_S e \nu}}\xspace}
\newcommand{\kzenu}{\mbox{\ensuremath{D_s \to K^0 e \nu}}\xspace}
\newcommand{\kstarenu}{\mbox{\ensuremath{D_s \to K^* e \nu}}\xspace}

\newcommand{\phinu}{\ensuremath{\phi e \nu}\xspace}
\newcommand{\etanu}{\ensuremath{\eta e \nu}\xspace}
\newcommand{\etapnu}{\ensuremath{\eta' e \nu}\xspace}
\newcommand{\fnu}{\ensuremath{f_0 e \nu}\xspace}
\newcommand{\ksnu}{\ensuremath{K_S e \nu}\xspace}
\newcommand{\kznu}{\ensuremath{K^0 e \nu}\xspace}
\newcommand{\kstarnu}{\ensuremath{K^* e \nu}\xspace}

\newcommand{\ds}{\ensuremath{D_s}\xspace}
\newcommand{\dsstar}{\ensuremath{D_s^*}\xspace}

\newcommand{\dsstards}{\ensuremath{D_s^*D_s}\xspace}

\newcommand{\fenupipi}{\mbox{\ensuremath{D_s \to f_0 e \nu, f_0 \to
      \pi\pi}}\xspace}

\newcommand{\kskch}{\ensuremath{K_S K^+}\xspace}
\newcommand{\kkpich}{\ensuremath{K^+ K^- \pi^+}\xspace}
\newcommand{\kskpizch}{\ensuremath{K_S K^+ \pi^0}\xspace}
\newcommand{\kskspich}{\ensuremath{K_S K_S \pi^+}\xspace}
\newcommand{\kkpipizch}{\ensuremath{K^+ K^- \pi^+ \pi^0}\xspace}
\newcommand{\kskppipich}{\ensuremath{K_S K^+ \pi^+ \pi^-}\xspace}
\newcommand{\kskmpipich}{\ensuremath{K_S K^- \pi^+ \pi^+}\xspace}
\newcommand{\pipipich}{\ensuremath{\pi^+ \pi^+ \pi^-}\xspace}
\newcommand{\pietach}{\ensuremath{\pi^+ \eta}\xspace}
\newcommand{\pipizetach}{\ensuremath{\pi^+ \pi^0 \eta}\xspace}
\newcommand{\pietapch}{\ensuremath{\pi^+ \eta', \eta' \to \pi^+ \pi^-
    \eta}\xspace} 
\newcommand{\pipizetapch}{\ensuremath{\pi^+ \pi^0
    \eta', \eta' \to \pi^+ \pi^- \eta}\xspace}
\newcommand{\pietaprhogamch}{\ensuremath{\pi^+ \eta', \eta' \to \rho^0
    \gamma}\xspace}

\newcommand{\ksk}{\ensuremath{K_S K}\xspace}
\newcommand{\kkpi}{\ensuremath{K K \pi}\xspace}

\newcommand{\kskspi}{\ensuremath{K_S K_S \pi}\xspace}
\newcommand{\kkpipiz}{\ensuremath{K K \pi \pi^0}\xspace}
\newcommand{\kskppipi}{\ensuremath{K_S K^+ \pi \pi}\xspace}
\newcommand{\kskmpipi}{\ensuremath{K_S K^- \pi \pi}\xspace}
\newcommand{\pipipi}{\ensuremath{\pi \pi \pi}\xspace}
\newcommand{\pieta}{\ensuremath{\pi \eta}\xspace}
\newcommand{\pipizeta}{\ensuremath{\pi \pi^0 \eta}\xspace}
\newcommand{\pietap}{\ensuremath{\pi \eta', \eta' \to \pi \pi
    \eta}\xspace} 

\newcommand{\pietaprhogam}{\ensuremath{\pi \eta', \eta' \to \rho^0
    \gamma}\xspace}

\newcolumntype{.}{D{.}{.}{-1}}
\newcolumntype{d}[1]{D{.}{.}{#1}}

\newcommand{\phienubr}{\ensuremath{2.14 \pm 0.17 \pm 0.09}}
\newcommand{\etaenubr}{\ensuremath{2.28 \pm 0.14 \pm 0.20}}
\newcommand{\etapenubr}{\ensuremath{0.68 \pm 0.15 \pm 0.06}}
\newcommand{\fenupipibr}{\ensuremath{0.13 \pm 0.02 \pm 0.01}}

\newcommand{\kzenubr}{\ensuremath{0.39 \pm 0.08 \pm 0.03}}
\newcommand{\kstarenubr}{\ensuremath{0.18 \pm 0.04 \pm 0.01}}

\begin{document}

\author{J. Hietala}\email{justinh@physics.umn.edu}
\author{D. Cronin-Hennessy}\email{hennessy@physics.umn.edu}
\affiliation{Department of Physics, University of Minnesota,
  Minneapolis, Minnesota 55455, USA}

\author{T. Pedlar} 
\affiliation{Department of Physics, Luther College,
  Decorah, Iowa 52101, USA}

\author{I. Shipsey} 
\affiliation{Sub-department of Particle Physics,
  Oxford OX1 3RH, United Kingdom}

\title{Exclusive \ds semileptonic
  branching fraction measurements}

\begin{abstract}
We measure absolute branching fractions for six exclusive \ds
semileptonic decays. We use data collected in the \mbox{CLEO-c}
detector from $e^+e^-$ annihilations delivered by the Cornell Electron
Storage Ring with a center-of-mass energy near 4170\mev. We
find \br(\phienu)~= (\phienubr)\%, \br(\etaenu)~= (\etaenubr)\%,
and \br(\etapenu)~= (\etapenubr)\% for the largest modes, where the
first uncertainties are statistical and the second are systematic. We
also obtain \br(\kzenu)~= (\kzenubr)\%, \br(\kstarenu)~=
(\kstarenubr)\%, and \br(\fenupipi)~= (\fenupipibr)\% for $f_0$ masses
within 60\mev of 980\mev. We use our results to determine the
$\eta-\eta'$ and $f_0$ mixing angles with $s\bar{s}$, and we combine
our results with lattice calculations to estimate $|V_{cs}|$. This
measurement improves upon the \ds semileptonic branching ratio
precision and provides a new approach for future work that eliminates
the \dsstar daughter photon reconstruction.
\end{abstract}


\maketitle


\section{INTRODUCTION}

\ds semileptonic decays have applications in both QCD tests and light
meson spectroscopy. Most notably, exclusive \ds decays to the dominant
modes (\phinu, \etanu, \etapnu) involve no light valence quarks and
thus provide an ideal opportunity for comparisons to lattice QCD
results~\cite{lattice_phi, lattice_eta}. Additionally, since the \ds
primarily couples to the final state hadron's $s\bar{s}$ component,
\ds decay rates can probe the quark content of
$\eta-\eta'$~\cite{donato, colangelo} and of the scalar
$f_0$~\cite{aliev_qcdsum, bediaga_f0phi, ke_qq} (including possible
glue components~\cite{etaetapglue, glue_sl}).

Further, inclusive semileptonic width measurements of strange and
non-strange $D$ mesons have revealed an interesting gap. The widths
for $D^\pm, D^0,$ and \ds decays should be equal in the Operator
Product Expansion (OPE), up to SU(3) symmetry breaking and
nonfactorizable components~\cite{voloshin_incl} (although phase space
considerations may not be trivial~\cite{rosner}). While the $D^\pm$
and $D^0$ inclusive widths are consistent with each other, the \ds
inclusive semileptonic width~\cite{ds_inclusive} falls some 16\%
lower, outside the range of experimental error. As the few lowest
lying resonances dominate $D^0$ and possibly $D^+$
semileptonics~\cite{dp_d0_exclusive, dp_d0_inclusive, pdg}, a higher
precision measurement of the analogous modes in \ds semileptonics
could shed light on this difference.

Although \ds exclusive semileptonic rates have been previously
studied~\cite{e687, cleo_old, focus}, the earlier measurements used
relative branching fractions and focused on only \phienu or $\ds \to
\phi \mu \nu$. These measurements are complicated by possible
interference between the reference mode, $\ds \to \phi\pi$, and other
$\ds \to KK\pi$ modes. BaBar~\cite{babar_kkenu} has more recently
obtained $\br(\phienu) = (2.61 \pm 0.03 \pm 0.17)\%$ in a relative
measurement using a 10 MeV mass requirement for $\phi \to KK$ and
taking $\ds \to KK\pi$ as their reference mode. In addition to its
inclusive \ds semileptonic measurement~\cite{ds_inclusive}, CLEO-c has
determined absolute branching fractions for six \ds exclusive
semileptonic modes in a partial ($310~\textrm{pb}^{-1}$) data
sample~\cite{cleo_exclusive} and performed another analysis for
\phienu and \fenu over a larger sample
($600~\textrm{pb}^{-1}$)~\cite{cleo_f0enu}. Our analysis improves upon
these results by using a novel technique that increases the efficiency
for all semileptonic modes and eliminates a limiting systematic in
prior measurments.

We use a data sample with an integrated $e^+e^-$ luminosity of
$586~\textrm{pb}^{-1}$ at a 4170 MeV center-of-mass energy, collected
in the CLEO-c detector~\cite{cleo_nim1, cleo_nim2}. The detector
provided both charged and neutral particle identification. Charged
particles followed a helical path through the detector's drift chamber
under the uniform 1.0 Tesla magnetic field, allowing particle
tracking, momentum determination, and mass identification from the
specific ionization ($dE/dx$). A Ring-Imaging Cherenkov detector
(RICH) improved charged particle identification for higher momentum
tracks, where $dE/dx$ does not give good separation. The RICH measured
the light cone given off by particles passing through a LiF radiator,
with an opening angle determined by the particle velocity. CLEO's CsI
electromagnetic calorimeter detected photons, measuring their energy
and direction. The calorimeter also contributed to identifying
electrons through $E/p$, the energy deposited by a charged particle in
the calorimeter relative to its momentum. Drift chamber tracks had a
momentum resolution of 0.35\% at 1 GeV, while calorimeter energy
measurements had a resolution of about 4\% at an energy of 100\mev and
about 2.2\% at an energy of 1\gev.~\cite{cleo_proposal}


\section{\dsstards EVENT IDENTIFICATION}

Most \ds production in electron-positron collisions at a 4170\mev
center-of-mass energy comes in the form of $\dsstards$ events with a
cross section of 0.92\nb, while $\ds\ds$ events make up another
0.03\nb~\cite{brian_scan}. By contrast, the cross section to other
charm events totals around 9\nb, with another 12\nb for $uds$
continuum. To cleanly separate candidate \ds events from other charm
and continuum, we completely reconstruct, or {\it tag}, one of the \ds
mesons in the event. We use 13 different \ds decay modes in our tag
reconstruction, listed in Table~\ref{tab:tag_counts}.

\begin{table}[!ht]
  \caption{\label{tab:tag_counts}Tag modes and counts. We list tag
    modes using their charges in $\ds^+$ decays for clarity, although
    the number of tags column contains the sum of results from both
    $\ds^+$ and $\ds^-$. The listed error is statistical.}
  \begin{center}
    \begin{tabular}{ l @{\hskip 0.5in} r @{ $\pm$ \hskip 0.05in } r } 
      \hline \hline
      \multicolumn{1}{ c }{$\ds^+$ mode} &
      \multicolumn{2}{c }{Number of tags} \\

      \hline

      \kskch & 6,226.7 & 101.2 \\
      \kkpich & 27,373.5 & 248.4 \\
      \kskpizch & 2,246.8 & 209.9 \\
      \kskspich & 1,125.5 & 76.5 \\
      \kkpipizch & 7,355.5 & 377.4 \\
      \kskppipich & 1,859.4 & 120.6 \\
      \kskmpipich & 3,377.3 & 100.0 \\
      \pipipich & 6,606.3 & 337.7 \\
      \pietach & 3,810.3 & 190.8 \\
      \pipizetach & 9,476.9 & 529.0 \\
      \pietapch & 2,386.6 & 65.6 \\
      \pipizetapch & 1,090.5 & 118.7 \\
      \pietaprhogamch & 4,272.3 & 193.3 \\
      \hline
      Sum & 77,207.5 & 880.2  \\ 
      \hline \hline
      \end{tabular}
    \end{center}
  \end{table}

The \dsstar decays to $\ds \gamma$ about 95\% of the time. The most
common state produced in \ds events then contains a $D_s^+$, a
$D_s^-$, and a photon. The standard approach would involve a tag
consisting of one \ds and the \dsstar daughter photon, leaving just the
other \ds. However, the \dsstar daughter photon reconstruction causes
both an efficiency loss (about 1/3 are lost) and a high fake rate
(about 50\% of the true total), with nontrivial systematic effects
given the accuracy of calorimeter simulations for low energy
deposition. We consequently do not reconstruct the \dsstar daughter
photon. This significantly improves our signal statistics and reduces
the problematic photon fakes, albeit at the expense of a clean
neutrino missing mass on the semileptonic side. Given the low
backgrounds from our \ds and electron selections, however, we see a
net improvement in our error by dropping the \dsstar daughter photon,
using only the reconstructed \ds as our tag, and constructing an
alternate method for signal determination (described in
Sec. \ref{sec:sl_reconstruction}).

Each tag mode's daughter particles have various track and shower
quality requirements to ensure proper \ds reconstruction. Each fitted
track must come within 5~mm of the interaction point in the radial
direction and within 5~cm in the beam direction. Each track must also
have at least 50\% of the expected drift chamber wire hits and fall
within the drift chamber's fiducial volume ($|\cos \theta| < 0.93$,
with $\theta$ measured from the beamline). Candidate pions are
required to have momenta above 50 MeV or 100 MeV (depending on the
mode's background) to avoid double counting by swapping soft pions
with the other side \ds. Candidate kaons must have a momentum above
125 MeV. Each track must have a $dE/dx$ consistent with its mass
hypothesis to within three standard deviations (3$\sigma$), and we add
a combined RICH and $dE/dx$ requirement for tracks in the RICH
fiducial region ($|\cos \theta| < 0.8$) when $dE/dx$ does not give
good separation (momenta above 700 MeV). Our photon candidates
(including $\pi^0$ and $\eta$ daughters) must have shower energies
above 30 MeV, and no tracks may lead to that shower.

Intermediate resonances receive additional selections. Our $\pi^0 \to
\gamma \gamma$ and $\eta \to \gamma \gamma$ candidates must have a
pull mass (standard deviation from nominal mass) within 3$\sigma$, and
the $\eta$ may not have both daughter showers detected in the
calorimeter's endcap region ($0.85 < |\cos \theta| < 0.93$). Candidate
$K_S$ must have a mass within 6.3 MeV (1.6$\sigma$) of their nominal
value. Our $\eta' \to \pi\pi\eta$ decays must involve a reconstructed
$\eta'$ mass within 10 MeV of its nominal value. The $\eta' \to
\rho^0\gamma$ mode has the wider mass requirement that the $\eta'$
mass falls between 920 MeV and 995 MeV, with a $\rho$ mass between 0.5
GeV and 1.0 GeV. We also require individual tag mode selections to
reject particular backgrounds. Specifically, no subset of particles
may form a $D^0$ or $D^\pm$ to avoid $D^*$ events (e.g. in
$KK\pi\pi^0$, the $KK\pi$ mass can not fall between 1860 MeV and 1880
MeV); two pions may not form a $K_S$ invariant mass except when
explicitly desired; and in the $\pi\pi\pi$ mode, treating a
reconstructed pion as a kaon can not form a $D^0$ mass with one of the
other pions.

As a final restriction on our \ds tag candidates, we ensure that they
have a momentum consistent with a \dsstards event through their recoil
mass. The \ds recoil mass is defined by
\begin{widetext}
\begin{equation}
M_\textrm{recoil} \equiv |p_\textrm{cm} - p_{\ds}| = \sqrt{\left
  (E_\textrm{cm} - \sqrt{|\mathbf{p_{\ds}}|^2 + M_{\ds}^2}\right )^2 -
  \left |\mathbf{p_\textrm{cm}} - \mathbf{p_{\ds}}\right|^2},
\end{equation}
\end{widetext}
where $p_\textrm{cm}, E_\textrm{cm},$ and $\mathbf{p_\textrm{cm}}$
correspond to the center-of-mass four vector, energy, and momentum,
respectively; $M_{\ds}$ is the nominal \ds mass; and
$\mathbf{p_{\ds}}$ denotes the reconstructed \ds momentum. The recoil
mass corresponds to the \dsstar mass for prompt \ds in \dsstards, and
it is fairly uncorrelated with the reconstructed invariant mass. We
require a minimum recoil mass of 2.051~GeV for \ksk, \kkpi,
\pieta, and \pietap; a minimum recoil mass of 2.101~GeV for
\pipipi; and a minimum recoil mass of 2.099~GeV for all other tag
modes. We only keep the best \ds candidate for each charge, as
determined by the recoil mass closest to the \dsstar mass
(2.112~GeV). This procedure successfully reconstructs around 7.2\% of
all prompt \ds decays and around 5.7\% of all secondary \ds decays
(those where the \ds came from a \dsstar, broadening their momentum
distribution).

\begin{figure*}[ht]
  \begin{center}
    \includegraphics[width=\textwidth]{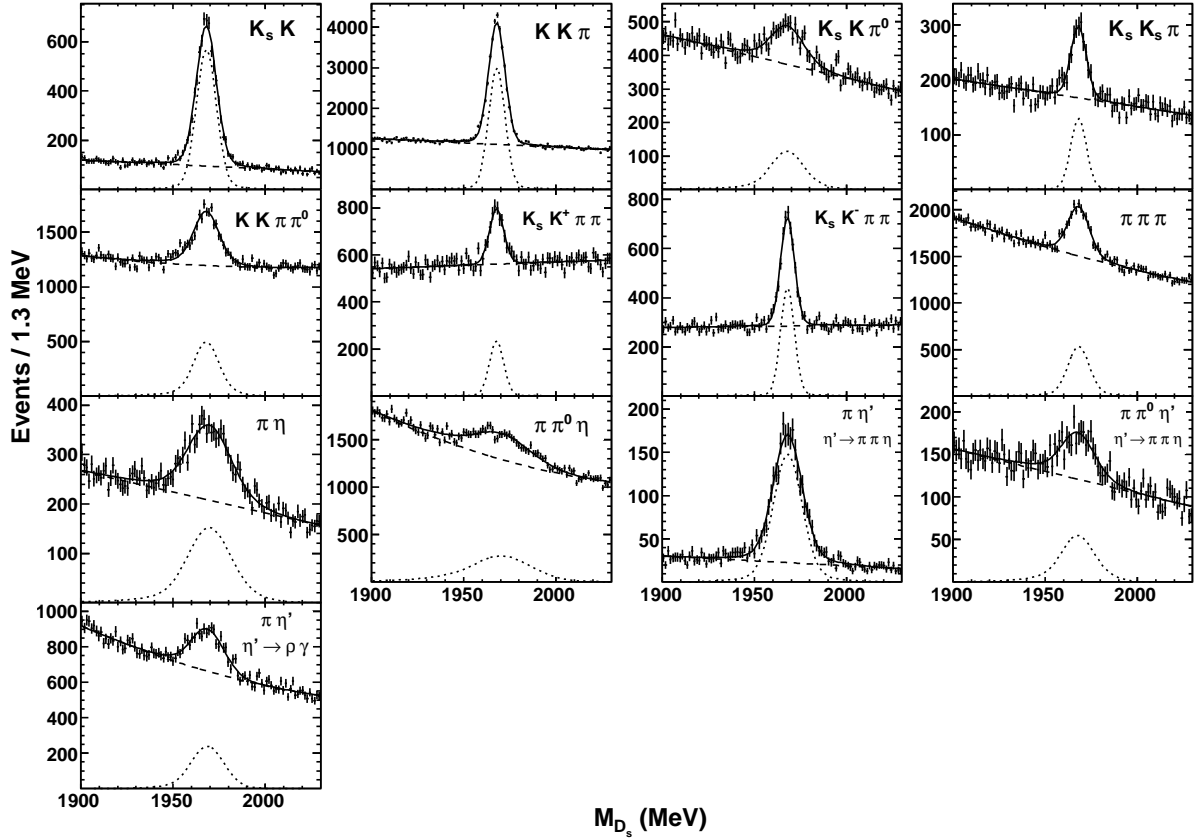}
    \caption{\label{fig:tag_inv_mass} Invariant \ds mass for each of
      our 13 tag modes. Points with error bars represent the data, the
      solid line represents our total fit, and the dotted and dashed
      lines give our signal and background fit components,
      respectively.}
  \end{center}
\end{figure*}

To obtain our total \ds tag counts, we fit the \ds invariant mass
spectrum for each tag mode, as shown in Figure
\ref{fig:tag_inv_mass}. We model our signal shape with either the sum
of two Gaussians (a double Gaussian) or a Gaussian added to another
with a power law tail (a \gcb~\cite{crystal_ball}). The tag modes
\ksk, \kkpi, \kskspi, \kskppipi, \kskmpipi, and \pietap each receive
the double Gaussian signal shape, while the other modes receive the
\gcb signal shape. We use a quadratic background for \kkpipiz,
\pipipi, \pipizeta, and \pietaprhogam, with a linear background for
the other tag modes. Table~\ref{tab:tag_counts} gives the tagged \ds
counts resulting from our fits.

\section{\label{sec:sl_reconstruction}SEMILEPTONIC RECONSTRUCTION}

Each semileptonic reconstruction involves an electron (positron)
identification. We use three parameters in a weighted combination to
identify a track as an electron. The most useful separation comes from
the energy deposited in the calorimeter relative to the particle's
momentum, $E/p$. We also include the particle's specific ionization in
the drift chamber ($dE/dx$) and RICH information. Our electron
efficiency varies by semileptonic mode but generally falls between
60\%--70\%, with most of the efficiency loss coming from a requirement
that the electrons have momenta above 200 MeV (above the pion and
electron $dE/dx$ crossing). Only 0.1\% of kaons in the appropriate
momentum range successfully fake an electron, while pions fake less
than 0.01\% of the time.

We also require that no semileptonic event have tracks from the
interaction point other than those accounted for in the tagged \ds,
the electron, and the semileptonic-side hadron. We considered a
similar constraint on extra energy in the calorimeter but did not find
it useful given the spurious showers that accompany hadronic
interactions.


Five of our six exclusive semileptonic measurements use a similar
technique. \phienu, \etapenu, \kzenu, \kstarenu, and \fenu all involve
finding the \ds tag, the semileptonic-side electron, and the
semileptonic-side hadron, then fitting the tagged \ds invariant mass
spectrum for the total number of semileptonic events. In these modes,
low backgrounds allow us to determine the event counts without
directly incorporating the semileptonic-side hadron's kinematic
information into the fit. \etaenu does see significant background from
photon fakes, so we instead perform a two-dimensional fit to the
tagged \ds invariant mass and the $\eta$ pull mass.

\subsection{\boldmath$\ds \to (\phi, \eta', K^0, K^*, f_0) e \nu$}

We reconstruct our semileptonic-side hadrons through the modes
$\phi \to KK$; $\eta' \to \pi\pi\eta, \eta \to
\gamma\gamma$; $K^0 \to K_S \to \pi\pi$; $K^* \to K\pi$; and $f_0 \to
\pi\pi$. We require the same daughter particle selections as for \ds tags,
with a few exceptions. Our $\phi \to KK$ decays produce soft kaons
that can decay in flight. Consequently, we remove the requirement that
the drift chamber has 50\% or more of the expected hits. We also do
not use the RICH information for kaons from a $\phi$. The $K^* \to
K\pi$ decay has a similar (but less severe) soft kaon problem, so we
relax its kaon hit requirement to 30\%. We apply a flight significance
selection in $K_S \to \pi\pi$ decays to ensure that the daughter pions
did not come from the interaction point ($\pi\pi$ vertex more than
4$\sigma$ from the interaction point). We also add a maximum flight
distance of 20~cm to avoid fake $K_S$ created near the
calorimeter. Given the low backgrounds, we implement loose mass
selections on our resonances: the reconstructed $\phi$ mass must be
within 15 MeV of the nominal mass on the low side and 30 MeV on the
high side ($-15 \mev < M_\phi^\textrm{recon} - M_\phi^\textrm{nom} <
30 \mev$), avoiding sensitivity to resonance effects near $KK$
threshold while retaining the high-side mass tail; the reconstructed
$\eta'$ mass must fall within 10\mev of its nominal value; $K_S$
follows the 6.3\mev mass cut listed with our tags; the $K^*$ mass must
be within 106\mev of its nominal value; and the $f_0$ mass must be
within 60\mev of 980\mev.

We see some background in our exclusive semileptonic modes from other
\ds semileptonic decays (e.g. $\fenu, f_0 \to KK$ background in
\phienu; $\phienu, \phi \to K_S K_L$ background in \kzenu). For
\phinu, we use our measured \fenu branching fraction and Monte Carlo
simulations with a Flatt\'{e} model~\cite{flatte76, flatte72} to
correct our observed branching fraction. In \kstarnu and \kznu, we cut
on the ``missing mass,'' which here corresponds to the invariant mass
of the neutrino and the \dsstar photon. This selection (mass squared
below $0.4\gev^2$ for \kznu and below $0.45\gev^2$ for \kstarnu)
distinguishes signal from background events with a missing
$K_L$. Finally, we ensure that we do not have \phinu, $\phi \to KK$
faking \kstarnu, $K^* \to K\pi$ by treating the $K^*$ pion as a kaon
and vetoing candidates with an invariant $KK$ mass less than
1.06\gev. We apply an explicit correction for remaining background
from other \ds semileptonic modes by using the background mode's
measured branching ratio and the efficiency with which it fakes the
target mode's selections. We additionally correct for the small number
of events (0.10--1.25, depending on semileptonic mode) with a true \ds
but a false hadron or non-semileptonic electron using Monte Carlo
predictions, cross-checked by data comparisons in the hadronic mass
sideband and alternate reconstructions for the electron.

\begin{figure*}[ht]
  \begin{center}
    \includegraphics[width=\textwidth]{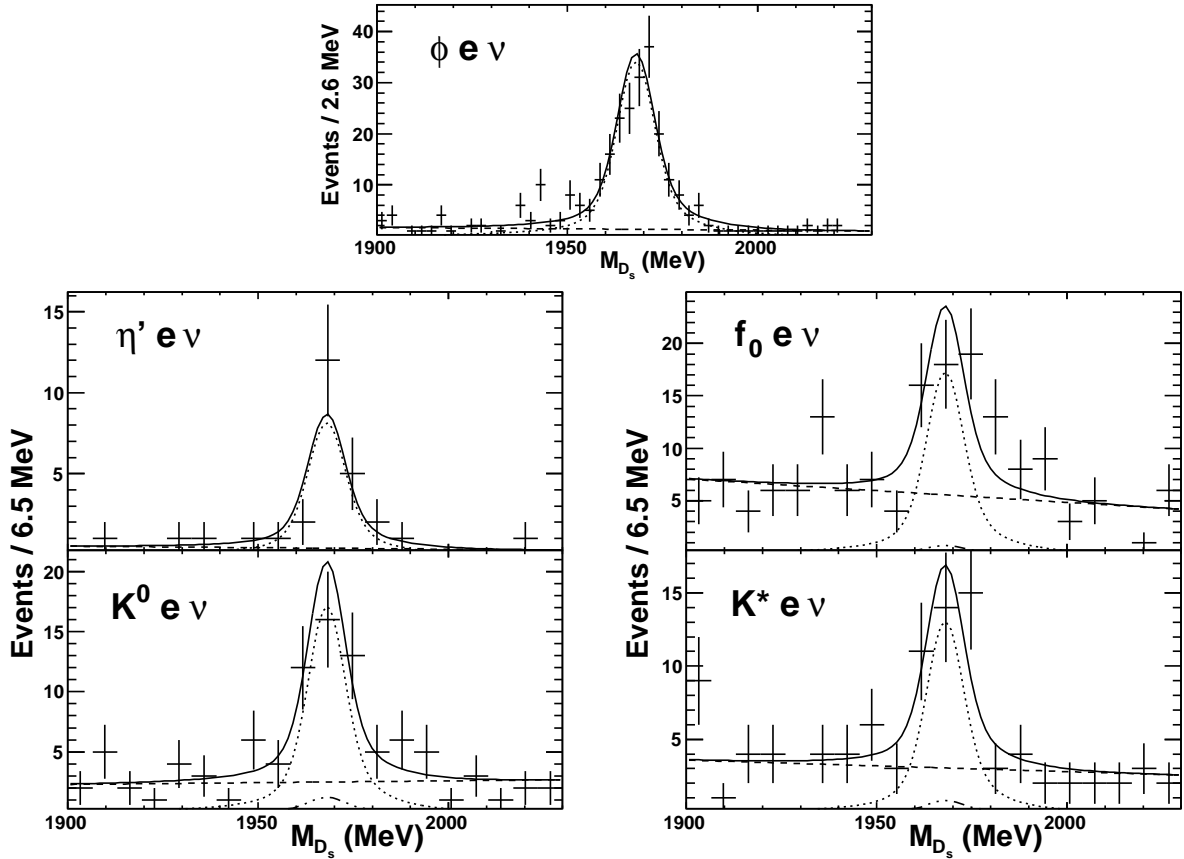}
    \caption{\label{fig:combined_henu} Tagged \ds invariant mass
      summed over all modes after semileptonic selections for \phinu,
      \etapnu, \fnu, \ksnu, and \kstarnu. Crosses represent data,
      while the solid line gives our total fit. The dotted line shows
      the signal part of our fit, the dashed line gives combinatoric
      background, and the small, peaking component represented by the
      dashed-dot line gives our background from other \ds semileptonic
      modes.}
  \end{center}
\end{figure*}

After finding an event with a valid \ds tag, electron, and
semileptonic-side hadron, we fit the tag's invariant mass. We take the
signal shape for each \ds tag mode from the results of that mode's
tagging fit. Each mode gets a linear or constant background based on
our Monte Carlo prediction for combinatoric background. We then
perform an unbinned, log likelihood fit on the data that is linked
across the 13 tag modes by a common branching ratio
constraint. Figure~\ref{fig:combined_henu} shows the results of our
fits, summed over all 13 tag modes.

\subsection{\boldmath\etaenu}

We reconstruct \etaenu through $\eta \to \gamma \gamma$. We use the
same selections as for $\eta$ in our \ds tags except for the pull mass
requirement, which we relax to 5$\sigma$ to give sufficient sideband
regions in our fits. After reconstructing the $\eta$, we also
implement a missing mass squared maximum of $0.5 \gev^2$ to avoid
backgrounds from other semileptonic modes that decay to $\eta$ (like
\etapenu, $\eta' \to \pi^0\pi^0\eta$).

We see several ``volunteer'' events in our \etaenu reconstruction,
where a true event gets reconstructed incorrectly. This happens when
the \dsstar daughter photon or a photon fake combines with a true
$\eta$ daughter photon to make a false $\eta$ combination, either in
addition to the true combination or as the only combination when the
true $\eta$ was missed. While the \dsstar daughter photon volunteer
rate can be determined from kinematics, the volunteer rate from fake
photon combinations depends upon detector effects that are not well
understood. We explicitly estimate the rate of these volunteer events
by reconstructing $D^0 \to K^*\eta$ in the much larger 3770 MeV CLEO-c
sample and incorporate the $\eta$ volunteer rate from that data's
result into our fits.

We then perform a two-dimensional fit to the reconstructed \ds tag
mass and the $\eta$ pull mass. As before, we use the results of our
\ds tagging fits to fix the \ds invariant mass shape. We take a signal
$\eta$ shape from the Monte Carlo with a single scale parameter. Both
the tag and the $\eta$ pull mass fits receive linear background
functions. We generate our two-dimensional fit function by multiplying
the signal and background tag functions by the signal and background
$\eta$ functions, taking separate normalizations for each background
mode and using a common branching ratio for the signal shapes across
each tag mode. We constrain our true \ds, false $\eta$ using our $D^0
\to K^*\eta$ study's volunteer rate, adjusted for the number of kaons
and pions in the \ds tag mode.

\begin{figure*}[ht]
  \begin{center}
    \includegraphics[width=\textwidth]{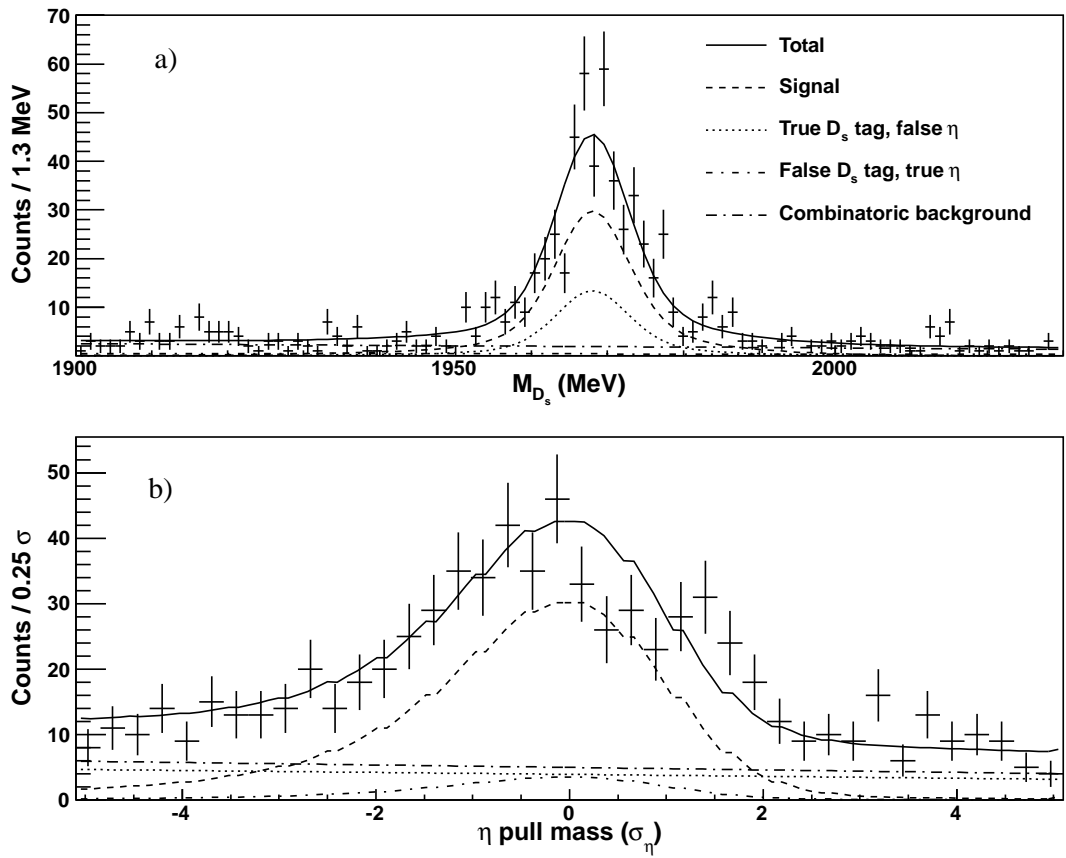}
    \caption{\label{fig:combined_etaenu} Our \etaenu reconstruction's
      two dimensional fit projections for (a) invariant \ds mass and
      (b) $\eta$ pull mass. Crosses represent data, while the lines
      show our total fit result and its four fit components.}
  \end{center}
\end{figure*}

Figure~\ref{fig:combined_etaenu} shows the \ds mass and $\eta$ pull
mass projections of our two-dimensional fits.

\subsection{Systematic Uncertainties}

Our dominant systematic errors (those with a relative error above
1\%) come from particle reconstruction, particularly from the soft
kaons frequently produced in \phienu and \kstarenu decays
(around 2\%); fit uncertainties on the \ds tag spectrum (2\%);
the effect of our Monte Carlo's form factor model on predicted
efficiencies (1\%--3\%); the choice of a best candidate for
the recoil mass (0\%--3\%); the mass resolution on our
$\eta'$, $K^*$, and $f_0$ selections (3\%); soft $K_S$
reconstruction in \ksenu (7\%); and $\eta$ reconstruction via
two photons in \etaenu and \etapenu (8\%).

We use $D^\pm \to K^\mp\pi^\pm\pi^\pm$ decays at 3770 MeV to estimate
the systematic error for charged kaon reconstruction, including
particle identification. We reconstruct a $D^\pm$ tag, then find an
additional $\pi^\mp\pi^\mp$. We fit the recoil mass spectrum for
events when we successfully reconstruct a kaon using our selections
and again for events when we did not reconstruct a kaon, giving us our
kaon efficiency. We perform this procedure for kaons of different
momenta (determined by the recoil momentum) and correct our Monte
Carlo efficiency in each momentum range accordingly.

We apply a similar approach for our $K_S$ reconstruction systematic,
although we need to use two modes to cover the full $K_S$ momentum
range: $\ds \to K_SK^\mp\pi^\pm\pi^\pm$ ($K^*K^*$) for lower momentum
$K_S$ and $\ds \to K_SK$ for higher momentum $K_S$. We again
reconstruct all particles but the $K_S$ (including the \dsstar
daughter photon), use the recoil momentum to determine the underlying
$K_s$ momentum region, and fit the recoil mass for found and not found
$K_S$ to determine the Monte Carlo efficiency in each $K_S$ momentum
range.

Our $\eta$ reconstruction systematic takes advantage of the relatively
high $\ds \to \pi\pi^0\eta$ rate, where we reconstruct the \ds tag,
the \dsstar daughter photon, and a $\pi\pi^0$ combination. To avoid
complications from the \dsstar photon resolution, we perform a
two-dimensional fit to the $\ds + \gamma$ recoil mass and the $\ds +
\gamma + \pi\pi^0$ recoil mass for our candidate events. We then do
another two-dimensional fit to the $\ds + \gamma + \pi\pi^0$ recoil
mass and $\eta$ pull mass for our succesfully reconstructed $\eta$
candidates. The ratio of these fits gives us our efficiency for $\eta$
reconstruction and the associated systematic error.

We determine the uncertainty on our \ds tag fits' signal shapes by
reconstructing analogous modes in the high-yield $D^\pm$ system and
adjusting the \ds fit functions' parameters to match the measured
$D^\pm$ mass resolutions. We estimate the systematic error on our \ds
tag fits' background shapes by using the Monte Carlo predicted
backgrounds in place of our linear or quadratic backgrounds.

To estimate the effects of an improper Monte Carlo mass resolution on
our $\eta'$, $K^*$, and $f_0$ intermediate resonances, we use the
reconstructed resolution from the clean modes $\ds \to \pi\eta'$, $\ds
\to K^*K$, and $\ds \to f_0\pi$, respectively. We generated Monte
Carlo using both the ISGW2 form factor model~\cite{isgw2} and a simple
pole model, then took the efficiency difference between the two as our
standard deviation for the semileptonic efficiency's systematic due to
uncertain form factors.


\section{RESULTS}

Table~\ref{tab:signal} gives the branching ratio results for each of
our six semileptonic modes, along with their efficiencies and number
of signal events. These results improve the existing precision by
about 20\% for the largest modes, \phinu and \etanu, and by 30\%--40\%
for the smaller branching fraction modes (other than $f_0$, which has
special considerations discussed below). The sum of our exclusive
modes has a branching fraction of $(5.80 \pm 0.27 \pm 0.30)\%$, which
falls below the inclusive rate of $(6.52 \pm 0.39 \pm 0.15)$\% by
$1.2\sigma$, possibly leaving a small role for semileptonic decays
with multiple hadrons.


\begin{table*}[!htbc]
  \caption{\label{tab:signal}Number of signal events, efficiencies
    (including all hadron branching fractions, like
    $\phi~\to~K^+K^-$), and final branching fractions for each
    semileptonic mode. We list our statistical error first, followed
    by our systematic error (combining both for the
    systematics-dominated efficiency error). Each mode uses $77,207.5
    \pm 880.2 \pm 1,675.4$ \ds tags. We have a soft correlation
    between the tags' systematic error and the systematic error on the
    number of signal events due to using a common \ds shape. We also
    have a moderate correlation in the systematic between semileptonic
    modes that is reflected in our sum's systematic error.}

  \begin{center}
    \begin{tabular}{l @{\hskip 0.5in} r @{ $\pm$ } r @{ $\pm$ } r 
        @{\hskip 0.5in} r @{ $\pm$ } r @{\hskip 0.5in} r @{ $\pm$ } r 
        @{ $\pm$ } r }
    \hline\hline

    Signal mode & \multicolumn{3}{c}{$N_\textrm{sig} \hspace{0.5in}$} 
    & \multicolumn{2}{c}{$\varepsilon_{\textrm{s}\ell} \hspace{0.5in}$} 
    & \multicolumn{3}{c}{\br (\%)} \\

    \hline

    \phienu & 206.7 & 16.4 & 2.3 & (12.5 & 0.5)\% & 2.14 & 0.17 & 0.09 \\ 
    \etaenu & 358.2 & 21.6 & 6.8 & (20.4 & 1.7)\% & 2.28 & 0.14 & 0.20 \\ 
    \etapenu & 20.1 & 4.4 & 0.3 & (3.8 & 0.4)\% & 0.68 & 0.15 & 0.06 \\ 
    \fenupipi & 41.9 & 7.8 & 0.6 & (21.2 & 1.0)\% & 0.13 & 0.03 & 0.01 \\ 
    \kzenu & 41.5 & 8.3 & 0.5 & (13.7 & 1.1)\% & 0.39 & 0.08 & 0.03 \\ 
    \kstarenu & 31.6 & 7.5 & 0.4 & (23.0 & 1.4)\% & 0.18 & 0.04 & 0.01 \\ 
\hline
    Sum &  \multicolumn{3}{c}{} & \multicolumn{2}{c}{} & 5.80 & 0.27 & 0.30 \\ 
    \hline\hline

    \end{tabular}
  \end{center}
\end{table*}


Table~\ref{tab:br_comparison} shows how this analysis's results
compare to prior results. Our \fenu, $f_0~\to~\pi\pi$ results give the
branching fraction for only $f_0~\to~\pi\pi$ that fall within a
$\pm$60~MeV mass window to avoid complications from the uncertain
$f_0$ width and the onset of nonlinear backgrounds at low $f_0$
masses. The previous analysis fit the $\pi\pi$ mass spectrum over a
wide range for their $f_0$ result. Both results are consistent if we
apply a $\pm$60~MeV mass requirement to their data as well.


\begin{table*}[!htbc]
  \caption{\label{tab:br_comparison}Most recent exclusive \ds
    semileptonic branching fraction measurements. Each of our modes is
    consistent with the previous CLEO-c measurements~\cite{cleo_f0enu,
      cleo_exclusive}, although we see an inconsistency in \phienu
    with BaBar's result~\cite{babar_kkenu}.}
  \begin{center}
    \begin{tabular}{l @{ \hskip 0.3in } r @{ $\pm$ } c @{ $\pm$ } c 
        @{ $\pm$ } l @{ \hskip 0.3in } r @{ $\pm$ } c @{ $\pm$ } l 
        @{ \hskip 0.3in } r @{ $\pm$ } c @{ $\pm$ } l}
    \hline\hline

    Signal mode & 
    \multicolumn{4}{c}{BaBar (\%)} &
    \multicolumn{3}{c}{CLEO-c (\%) $\hspace{0.2in}$} & 
    \multicolumn{3}{c}{This analysis (\%)} \\

    \hline

    \phienu & 
       2.61 & 0.03 & 0.08 & 0.15 &
       2.36 & 0.23 & 0.13 &
       2.14 & 0.17 & 0.09 \\
    \etaenu & 
       \multicolumn{4}{c}{--- $\hspace{0.3in}$} &
       2.48 & 0.29 & 0.13 &
       2.28 & 0.14 & 0.20 \\
    \etapenu & 
       \multicolumn{4}{c}{--- $\hspace{0.3in}$} &
       0.91 & 0.33 & 0.05 &
       0.68 & 0.15 & 0.06 \\
    \fenupipi & 
       \multicolumn{4}{c}{Seen $\hspace{0.3in}$} &
       0.20 & 0.03 & 0.01 &
       0.13 & 0.03 & 0.01 \\
    \ksenu & 
       \multicolumn{4}{c}{--- $\hspace{0.3in}$} &
       0.19 & 0.05 & 0.01 &
       0.20 & 0.04 & 0.02 \\
    \kstarenu & 
       \multicolumn{4}{c}{--- $\hspace{0.3in}$} &
       0.18 & 0.07 & 0.01 &
       0.18 & 0.04 & 0.01 \\
    \hline\hline

    \end{tabular}
  \end{center}
\end{table*}



\section{DISCUSSION}

Various theoretical predictions have been made for relative or
absolute \ds semileptonic decay rates~\cite{isgw2, lattice_phi,
  colangelo, rosner, bediaga_f0phi, ke_qq, aliev_qcdsum, melikhov,
  wei, azizi, colangelo_bs, khlopov}. Some predictions combine with
our measured results to determine meson mixing angles. For instance,
if we take $\eta$ and $\eta'$ to be purely $q\bar{q}$ states, the \ds
semileptonic decays to $\eta$ and $\eta'$ can extract the $\eta-\eta'$
mixing angle. For the mixing angle defined by
\begin{equation}
\begin{split}
\Ket{\eta'} &= \sin \phi \Ket{n\bar{n}} + \cos \phi \Ket{s\bar{s}} \\
\Ket{\eta} &= \cos \phi \Ket{n\bar{n}} - \sin \phi \Ket{s\bar{s}}
\end{split}
\end{equation}
with $\Ket{n\bar{n}} = \frac{1}{\sqrt{2}}\Ket{u\bar{u} + d\bar{d}}$,
the ratio of semileptonic widths gives~\cite{donato}
\begin{equation}
\frac{\Gamma(\etapenu)}{\Gamma(\etaenu)} = R_D \cot^2 \phi,
\end{equation}
where $R_D$ contains the relative phase space and the ratio of
integrated form factors. Anisovich, et al.~\cite{etaetapglue} have
used a monopole quark transition form factor to estimate $R_D~=~0.23$,
which combines with our result to give an $\eta-\eta'$ mixing angle of
$\phi~= \ang{41}~\pm \ang{4}$. If the constituent quark transition
form factor ratio is instead taken to be unity, $R_D~=~0.28$ and we
get $\phi~= \ang{44}~\pm \ang{4}$.

We can compare these results to the $SU(3)$ mixing angle given by
\begin{equation}
\begin{split}
\Ket{\eta'} &= \sin \theta \Ket{\eta_0} + \cos \theta \Ket{\eta_8} \\
\Ket{\eta} &= \cos \theta \Ket{\eta_0} - \sin \theta \Ket{\eta_8},
\end{split}
\end{equation}
where the singlet and octet states follow $\Ket{\eta_0} =
\frac{1}{\sqrt{3}}\Ket{u\bar{u} + d\bar{d} + s\bar{s}}$ and
$\Ket{\eta_8} = \frac{1}{\sqrt{6}}\Ket{u\bar{u} + d\bar{d} -
  2s\bar{s}}$. The bases relate to each other through $\theta = \phi -
\arctan \sqrt{2}$, with $\theta = 0$ corresponding to $SU(3)$
symmetry. In the $SU(3)$ basis, our results become $\theta = \ang{-13}
\pm \ang{4}$ with the monopole form factor and $\theta = \ang{-11} \pm
\ang{4}$ for the flat form factor.

Alternately, the assumption of an $\eta'$ state consisting of only
$q\bar{q}$ can be loosened by allowing for a glue component. In this
case, we can use $D^+$ semileptonic decays to cancel the glue
component through the ratio~\cite{donato}
\begin{equation}
\frac{\Gamma(\etapenu) / \Gamma(\etaenu)}{\Gamma(D^+ \to \eta' e \nu)
  / \Gamma(D^+ \to \eta e \nu)} = \cot^4 \phi.
\end{equation}
Here, the phase space and form factor ratio $R_D$ is assumed to be the
same for $D^+$ and $D_s$ decays. Combining our \ds results with the
$D^+$ data~\cite{Yelton:2010js} gives
$\phi~=~\ang{42}~\pm~\ang{2}~\pm~\ang{2}$
($\theta~=\ang{-13}~\pm~\ang{2}~\pm~\ang{2}$), where the first error
comes from the $D^+$ measurement and the second comes from our
measurement.

The $f_0$ mixing angle may also be extracted by comparisons to
theoretical calculations. Several such estimates of the $f_0$ decay
rate exist~\cite{aliev_qcdsum, bediaga_f0phi, ke_qq}, which
collectively set the branching fraction at $\br(\fenu) =
(0.41$--$0.55)\% \times \cos^2 \theta$. We use a Flatt\'{e} model with
a $\Gamma_{f_0}$ range from $50\mev$--$100\mev$, an $M_{f_0}$ range
from $970\mev$--$990\mev$, and $\Gamma(f_0 \to K^+K^-)/\Gamma(f_0 \to
\pi^+\pi^-)$ values taken from experiment~\cite{babarf0, bes2f0} to
estimate the fraction of $f_0 \to \pi^+\pi^-$ in our $\pm$60\mev
window. These combine with our \fenu measurement to yield an
$s\bar{s}$ mixing angle of $\cos^2 \theta = 0.94 \pm 0.26 \pm 0.07 \pm
0.19$, where the first error comes from the range of predictions, the
second error comes from the uncertain $f_0$ mass and width, and the
third error comes from our measurement. Ignoring the nonphysical range
and treating the errors as independent gives a mixing angle of
$\theta~=~\ang{20}^{+\ang{32}}_{\ang{-20}}$.

Additionally, combining our \phienu measurement with lattice
calculations determines a $\left|V_{cs}\right|$
value~\cite{lattice_phi}. We use~\cite{lattice_communication}
\begin{equation}
\frac{\br(\phienu)}{\left|V_{cs}\right|^2} = (2.52 \pm 0.22 \pm
0.15)\%, 
\end{equation}
where the first error comes from the \phienu lattice simulation, and
the second error comes from complications due to the strong $\phi \to
KK$ decay (not a ``gold-plated'' decay). This yields
$\left|V_{cs}\right| = 0.921 \pm 0.041 \pm 0.049$, with our
measurement uncertainty generating the first error and the combination
of both lattice uncertainties giving the second error. The
$\left|V_{cs}\right|$ result falls within one standard deviation of
the best current value ($0.986 \pm 0.016$)~\cite{pdg}.


\section{CONCLUSION}

We have used CLEO-c's 4170\mev data to measure semileptonic decays for
the six exclusive modes $\ds \to (\phi, \eta, \eta', f_0, K^0, K^*) e
\nu$. Our procedure uses additional data for four modes (\etanu,
\etapnu, \kznu, and \kstarnu) and involves a new technique in which
the \dsstar daughter photon does not get reconstructed, significantly
increasing the available statistics. We see \br(\phienu)~=
(\phienubr)\%; \br(\etaenu)~= (\etaenubr)\%; \br(\etapenu)~=
(\etapenubr)\%; \br(\kzenu)~= (\kzenubr)\%; \br(\kstarenu)~=
(\kstarenubr)\%; and \br(\fenupipi)~= (\fenupipibr)\% within 60\mev of
the $f_0$ mass. Our measurements show that these six exclusive modes
nearly saturate the inclusive \ds width.

We also combined our results with theoretical predictions and other
measurements to extract an $\eta-\eta'$ mixing angle of
$\phi~=~\ang{42}~\pm~\ang{2}~\pm~\ang{2}$ and an $f_0$ mixing angle
with $s\bar{s}$ of $\theta~=~\ang{20}^{+\ang{32}}_{\ang{-20}}$.


\begin{acknowledgments}

We would like to thank M.B. Voloshin for useful discussions on $D$
meson semileptonics, particularly inclusive decays. We thank
G.P. Lepage for sharing some of his lattice knowledge wtih us. This
analysis uses CLEO-c data; as members of the retired CLEO
collaboration, we appreciate the afforded opportunity to revive it one
more time. We also gratefully acknowledge the CESR staff for their
efforts to provide the good run conditions and luminosity needed for
this and all other CLEO work over four decades.

\end{acknowledgments}


%

\newpage

\end{document}